\documentclass[letterpaper]{amsart}
\usepackage[leqno]{amsmath}
\usepackage{amsfonts,amssymb,amsthm,amscd}

\usepackage{eucal}

\newtheorem{theorem}{Theorem}[section]
\newtheorem{proposition}[theorem]{Proposition}
\newtheorem{lemma}[theorem]{Lemma}
\newtheorem{corollary}[theorem]{Corollary}
\newtheorem{assumption-definition}[theorem]{Assumption-Definition}
\theoremstyle{definition}
\newtheorem{definition}[theorem]{Definition}
\newtheorem{example}[theorem]{Example}
\theoremstyle{remark} \newtheorem{remark}[theorem]{Remark}
\theoremstyle{definition} 
\numberwithin{equation}{section}

\newcommand{\field}[1]{\ensuremath{\mathbb{#1}}}
\newcommand{\CC}{\field{C}}

\newcommand{\ZZ}{\field{Z}}

\newcommand{\complex}[1]{\mathsf{#1}}

 \newcommand{\CCC}{\complex{C}}

\newcommand{\lie}[1]{\ensuremath{\mathfrak{#1}}}
\newcommand{\g}{\lie{g}}

\newcommand{\cover}[1]{\mathcal{#1}}

\newcommand{\sheaf}[1]{\underline{\mathnormal{#1}}}

 \DeclareMathOperator{\Tot}{Tot}

 \DeclareMathOperator{\ad}{ad}

\newcommand{\HHH}{\mathbb{H}}

\newcommand{\deltacheck}{\Check\delta} 

\newcommand{\var}{\boldsymbol{\delta}} 

\newcommand{\abs}[1]{\lvert#1\rvert}

\newcommand{\eqdef}{\overset{\mathrm{def}}{=}}

\begin{document}
\title{%
  Homological algebra of multivalued action functionals }
\author{%
  Ettore Aldrovandi }
\address{%
  Department of Mathematics\\
  Florida State University\\
  Tallahassee, FL 32306-4510, USA }
\email{%
  aldrovandi@math.fsu.edu }

\begin{abstract}
  We outline a cohomological treatment for multivalued
  (classical) action functionals. We point out that an
  application of Takens' theorem, after Zuckerman, Deligne and
  Freed, allows to conclude that multivalued functionals yield
  globally defined variational equations.
\end{abstract}

\maketitle

\section{Introduction, main definitions and statement of the results}
\label{sec:intr-stat-results}

\subsection{General motivation}
\label{sec:general}

In recent years there has been a continuously growing interest in
the analysis of multivalued action functionals. The precise
meaning of ``multivalued functional'' will be defined below; for
now, we can heuristically define an action functional to be
``multivalued'' if it is given in terms of a collection of
Lagrangian densities on the manifold $M$ of parameters (the
``space-time'') which do \emph{not} glue into a globally defined
differential form (of top degree on $M$).

Amongst the primary motivations to study multivalued actions are
the need to incorporate topological
terms~\cite{gaw1988,freed:hialg,alvarez1985}, and the emergence
of dynamical fields of new geometric content, such as curvings
and connective structures on Gerbes with abelian band, like the
$B$-field in String Theory, see e.g.~\cite{freedwitten:anom}, and
more recently, differential cohomology~\cite{freed:diff-coho}.
Multivalued actions can also arise in other geometric contexts,
typically when a Lagrangian density, that is, a top-degree form,
is produced from a local ansatz, where ``local'' means that the
construction leading to the Lagrangian density is carried out
with respect to an explicit choice of an open neighborhood or
chart $U\rightarrow M$. Such is the case, for example, for the
Liouville action constructed in~\cite{zogtak1987-2} to
investigate the Weil-Petersson form on the Teichm\"uller space of
compact Riemann surfaces of genus $g$, and its chiral ``half''
used in~\cite{aldtak1997,aldtak2000} that yields a variational
characterization of the universal projective family. It seems
that in all the examples a proper definition of the action
usually leads to a generalization of the Lagrangian density as a
cocycle in a \v{C}ech resolution with respect to a chosen open
cover of $M$.  The \v{C}ech paradigm is in fact a fundamental one
when constructing action functionals. In our work with
L.A.~Takhtajan (cf.~\cite{aldtak2000}) we stressed its universal
nature with respect to the choice of the cover, emphasizing the
need (and the possibility) to work with arbitrary coverings. In
this way we put the accent on the cocycle itself, rather than on
the class it represents: indeed the former is what is usually
explicitly computable, given a choice of a covering of $M$.

In geometric applications, such as in~\cite{zogtak1987-2}
or~\cite{aldtak2000}, we need to work with the first and higher
variations of the relevant functionals, thus we work firmly
within the context of Classical Field
Theory~\cite{delfreed1999-2}. It is obviously of primary
importance to ensure that the variational principle associated to
these multivalued actions yield well defined variational
problems. In other words, we need to ensure that the resulting
Euler-Lagrange equations be globally defined on $M$.

The aim of this note is to point out a simple mechanism by which
these (classical) multivalued action functionals yield a globally
defined variational equation. Assuming the variations of the
relevant dynamical fields glue appropriately on $M$, we show that
Takens' results on the variational bicomplex~\cite{takens1981}
(see also~\cite{delfreed1999-2,zuck:action}) force the local
variations subordinate to a covering of $M$ to glue into a global
one.

\subsection{Multivalued functionals}
\label{sec:mult-funct}

The meaning of ``multivalued functional'' in this context is as
follows. Let $M$ be a manifold of dimension $n$, assumed to be
compact for simplicity, and let $\cover{U}_M = \{ U\}_{i \in I}$
be an open cover. For any $p$ let $\sheaf{A}_M^p$ be the sheaf of
$p$-forms, with $A^p(M)$ the corresponding module of global
sections. (We assume the smooth forms to be $\CC$-valued, in
general.) Consider the datum of a smooth $n$-form $\omega^{(0)}_i
\in \sheaf{A}^n_M(U_i)$ for each $U_i$. Each $\omega^{(0)}_i$ is
interpreted as a Lagrangian ``density'' and if $x^1_i, \dots ,
x^n_i$ are local coordinates on $U_i$, we write $\omega^{(0)}_i =
L_i dx^1_i\wedge \dots \wedge dx^n_i$. We assume the smooth
function $L_i$ (the ``Lagrangian'') depends on a section of some
fiber bundle $E\overset{\pi}{\rightarrow} M$.  (In fact this can
be generalized to the situation where we have submersions $E_i
\rightarrow U_i$ satisfying reasonable descent conditions, see
below.) We need to compare two local Lagrangian densities
$\omega^{(0)}_i$ and $\omega^{(0)}_j$ on $U_{ij} = U_i \cap U_i$.
We consider the following two possibilities:
\begin{enumerate}
\item $\omega^{(0)}_i = \omega^{(0)}_j$ for any $i,j\in I$, that
  is the local lagrangian densities glue to form a globally
  defined $n$-form $\omega^{(0)}$ on $M$. In particular we can
  define an action functional by integration over $M$:
  \begin{displaymath}
    S = \langle [\omega^{(0)}],[M] \rangle = \int_{M}\omega^{(0)}\,.
  \end{displaymath}
  As a result, the corresponding variational principle will be
  well defined on $M$.
\item 
  \begin{math}
    \omega^{(0)}_j - \omega^{(0)}_i = d\omega^{(1)}_{ij}\,,
  \end{math}
  where $\omega^{(1)}_{ij}\in \sheaf{A}_M^{n-1}(U_{ij})$ is a
  smooth $(n-1)$-form. We refer to this case as ``multivalued'',
  owing to the non-uniqueness of the Lagrangian density.
\end{enumerate}
In the second case above, the procedure to construct an action
functional is by now standard. We consider the \v{C}ech-de Rham
complex~\cite{bott1982} $\check{C}^q(\cover{U}_M,\sheaf{A}_M^p)$
relative to the cover $\cover{U}_M$. If we assume this cover to
be good, then we can construct forms
$\omega^{(0)},\omega^{(1)},\dots ,\omega^{(n)}$, with
$\omega^{(q)}$ an $(n-q)$-form on a $q$-fold intersection, from
the descent relation
\begin{equation}\label{eq:2}
  (-1)^{n-q} \deltacheck \omega^{(q)} = d \omega^{(q+1)}\,,
\end{equation}
where $\deltacheck$ is the \v{C}ech coboundary. We obviously have
$d \omega^{(0)} = 0$ for dimensional reasons, and we can close
the descent condition at the last step, namely $\deltacheck
\omega^{(n)}=0$, by invoking the fact that $H^{n+1}(M^n,\CC) =
0$. In this way the sequence of forms
$\omega^{(0)},\omega^{(1)},\dots ,\omega^{(n)}$ determines a
total cocycle $\Omega$ of degree $n$ in the single complex
associated to the \v{C}ech-de Rham one with total differential $D
= d\pm \deltacheck$, namely we have $D\Omega = 0$. Since the
\v{C}ech-de Rham complex computes $\HHH^\bullet(M,
\sheaf{A}_M^\bullet) \cong H^\bullet (M,\CC)$, the cocycle
$\Omega$ represents a class of degree $n$ which can then be
evaluated against the fundamental class of $M$. More precisely,
$\Omega$ can be evaluated against a representative $\Sigma$ of
$[M]$ and the resulting number
\begin{equation}
  \label{eq:1}
  S = \langle \Omega , \Sigma\rangle \equiv \langle [\Omega],
  [M]\rangle 
\end{equation}
can be taken to be the action functional determined by the
collection of local Lagrangian densities $\{ \omega^{(0)}_i\}$.

We slightly modify this setup: in many actual examples the
descent equations are simply manifestly satisfied up to the last
step (so the assumption that the cover is good is not really
used), where we may have $\deltacheck \omega^{(n)} = c\in \ZZ$,
instead of simply $\deltacheck \omega^{(n)}= 0$. To accommodate
this fact it is worth working with Deligne cohomology, as pointed
out in~\cite{gaw1988}, namely we replace the de Rham complex
$\sheaf{A}_M^\bullet$ with the augmented one $\ZZ_M \rightarrow
\sheaf{A}_M^\bullet$, where the first arrow is just the
inclusion~\cite{delfreed1999-2}. Thus $\Omega$ will be
interpreted as a total cocycle of degree $n+1$ corresponding to
the sequence $\omega^{(0)},\omega^{(1)},\dots ,\omega^{(n)}, c$
in the \v{C}ech resolution of the Deligne complex above. If
indeed the cover is good, the \v{C}ech resolution computes the
\emph{Deligne cohomology} group $\HHH^{n+1}(M, \ZZ_M \rightarrow
\sheaf{A}_M^\bullet) \eqdef H_\mathcal{D}^{n+1}(M,\ZZ)\cong
H^n(M,\CC^*)$. Therefore the result of~\eqref{eq:1} is to be
interpreted as the \emph{exponential} of the action (written
additively), rather than the action itself. In any event, the
multivalued functionals we have in mind are precisely those
\v{C}ech cocycles arising from a collections of local lagrangian
densities in the manner we have just explained.

\subsection{Main definitions and results}
\label{sec:Main-defin-results}

Let $M$, $\cover{U}_M$, $\sheaf{A}_M^\bullet$ as above.  Let
$\ZZ_\mathcal{D}^\bullet : \ZZ_M \rightarrow \sheaf{A}_M^0
\rightarrow \sheaf{A}_M^1 \rightarrow \dotsb $ be the
\emph{augmented} de Rham complex---the Deligne complex of length
$n+1$. Let $\CCC^{p,q} = \check{C}^q(\cover{U}_M,
\ZZ_\mathcal{D}^p)$ the bicomplex with differentials, $d$ and
$\deltacheck$. The associated simple complex will have total
differential $D = d +(-1)^p\deltacheck$ acting on the homogeneous
components $\CCC^{p,\bullet}$. (Note that in degree zero $d$ is
just the inclusion $\ZZ_M$ into the smooth functions
$\sheaf{A}_M^0$.) Let $\{ \omega^{(0)}_i \}$ be a collection of
lagrangian densities subordinated to the open cover
$\cover{U}_M$. (In other words, a $0$-cochain on $\cover{U}_M$
with values in $\sheaf{A}_M^n$.)  For the purpose of this
introduction we will assume each $\omega^{(0)}_i$ depends (in a
local way, see below) on the restriction to $U_i$ of a section
$\phi$ of a smooth bundle $E\overset{\pi}{\rightarrow} M$. (We
omit to display this dependency in the notation.)
\begin{definition}
  \label{def:1}
  A \emph{multivalued Lagrangian cocycle} is a cocycle $\Omega =
  \omega^{(0)} + \omega^{(1)} + \dots + \omega^{(n)}+ c$ of total
  degree $n+1$ in the total simple complex associated to
  $\CCC^{p,q}$. The homogeneous members satisfy the descent
  condition~\eqref{eq:2} plus the relation $\deltacheck
  \omega^{(n)} = c$. If $\Sigma$ represents the fundamental class
  $[M]$ of $M$, the \emph{multivalued action functional}
  associated to $\Omega$ is given by the evaluation~\eqref{eq:1}.
\end{definition}
For each member $U_i$ of the cover $\cover{U}_M$, we consider the
variational bicomplex on $\mathcal{S}_i \times U_i$, where
$\mathcal{S}_i$ is the restriction to $U_i$ of the space of
smooth global sections of $E$. We have \emph{two} exterior
differentials: $\var$ in the field direction, and $d$ has been
already introduced. For simplicity we assume that $\var d = d
\var$, and include an explicit sign for the total differential
$\var \pm d$.

The variation $\var\omega^{(0)}$ can be written as
\begin{equation}
  \label{eq:3}
  \var \omega^{(0)}_i = a^{(0)}_i + d\gamma^{(0)}_i\,.
\end{equation}
It follows from Takens' theorem~\ref{thm:2} (see
\cite{takens1981}, and also \cite{delfreed1999-2}) that the
decomposition is unique (see below), where the variation
$a^{(0)}_i$---the non exact part of $\var\omega^{(0)}$---is a
\emph{source form} of degree $(1,n)$ in the variational
bicomplex, and $\gamma^{(0)}_i$ has degree $(1,n-1)$. The source
form $a^{(0)}_i$ determines a differential equation which
coincides with the classical Euler-Lagrange equations once a
coordinate system has been chosen. We call $\mathcal{M}_i$ the
zero-locus in $\mathcal{S}_i$ defined by the Euler-Lagrange
equation.  Following \cite{delfreed1999-2}, we call
$\gamma^{(0)}_i$ the Cartan form.

Then the main result is
\begin{theorem}
  \label{thm:1}
  The cochain $\{ a^{(0)}_i \}_{i\in I}$ is a $0$-cocycle, that
  is $a^{(0)}_i = a^{(0)}_j$, so it defines a globally defined
  $(1,n)$-source form $a^{(0)}$. The variation of the total
  cocycle $\Omega = \omega^{(0)} + \omega^{(1)} + \dots +
  \omega^{(n)}+ c$ is solely due to the source form up to a total
  coboundary, namely we have
  \begin{equation}
    \label{eq:4}
    \var\Omega = a^{(0)} + D\Gamma\,,
  \end{equation}
  where $\Gamma \equiv \sum_{q=0}^{n-1}\gamma^{(q)}$ is a chain
  of total degree $n$ in the total complex of
  $\CCC^{\bullet,\bullet}$ such the last component
  $\gamma^{(n)}=0$.
\end{theorem}

$\Gamma$ may be called the \emph{global Cartan form} associated
to $\Omega$. Notice that $\Gamma$ is really (up to an index
shift) an object in the genuine \v{C}ech-de Rham complex.

A number of corollaries are almost immediately available. First,
we obviously have
\begin{corollary}
  The zero-loci $\mathcal{M}_i\subset \mathcal{S}_i$ determined
  by the Euler-Lagrange equations relative to
  $\{a^{(0)}_i\}_{i\in I}$ glue into a global locus
  $\mathcal{M}$.
\end{corollary}
\begin{remark}
  Note that the statement of the corollary is not completely
  vacuous. In light of the assumptions in~\ref{def:assumption},
  it means that regardless of the nature of the allowed fields,
  the loci $\mathcal{M}_i$ will always describe global geometric
  objects on $M$.
\end{remark}
Furthermore, in ref. \cite{zuck:action} Zuckerman introduces the
variational differential of the Cartan form and calls it the
\emph{universal conserved current}. Thus for each $i\in I$ we
would consider the local universal conserved current
$\theta^{(0)}_i = \var\gamma^{(0)}_i$, which is a form of
bidegree $(2,n)$ in the variational complex.  Elementary
manipulations show that the main property of the Cartan form is
that
\begin{equation}
  \label{eq:5}
  \var \theta^{(0)}_i = 0\,,\qquad d \theta^{(0)}_i = -\var
  a^{(0)}_i\,,
\end{equation}
It follows from \eqref{eq:5} that the restriction of the
universal current to $\mathcal{M}_i$ is a \emph{closed}
$(2,n)$-form (in the variational bicomplex). To complete the
picture, we consider the global current $\Theta \eqdef
\var\Gamma$ and similarly to ref.~\cite{zuck:action} we have:
\begin{proposition}
  \label{prop:1}
  The global current $\Theta$ satisfies
  \begin{equation}
    \label{eq:6}
    \var \Theta = 0\,,\qquad D \Theta = -\var a^{(0)}\,
  \end{equation}
  so it is a conserved current. The restriction to $\mathcal{M}$
  is closed with respect to $\var +D$.
\end{proposition}
In analogy with the variation being a homogeneous element in the
\v{C}ech resolution of $\ZZ_\mathcal{D}^\bullet$, thanks to
Theorem~\ref{thm:1}, the proposition shows the $D$-differential
(and obviously the total differential) of the global current
$\Theta$ is also a homogeneous object of pure bidegree $(n+1,0)$
in the \v{C}ech-Deligne complex. 

\subsection{Organization}
\label{sec:Organization}
This note is organized as follows. In the first part of
sect.~\ref{sec:Setup} we collect some notation and some notions
we need about Deligne complexes and variational bicomplexes. In
subsect.~\ref{sec:Dynamical-fields} we make more precise
assumptions on the allowed objects in the variational process (in
particular relaxing the conditions stated in
subsect.~\ref{sec:Main-defin-results}) and we state a lemma on
the gluing properties of the resulting local variational
complexes. Once this is done, the proofs of Thm.~\ref{thm:1} and
Prop.~\ref{prop:1} reduce to a homological manipulation of
various differential complexes. They are presented in some detail
in sect.~\ref{sec:Proofs}. Finally, we draw some conclusions and
look at possible future directions in
sect.~\ref{sec:Conclusions}.

\section{Setup}
\label{sec:Setup}

We keep the assumptions on $M$, $\cover{U}_M$,
$\sheaf{A}_M^\bullet$ made in sect.~\ref{sec:intr-stat-results}.
Also, we use the notation $U_{ij}$ for $U_i\cap U_j$ and $U_{ijk} 
= U_i\cap U_j\cap U_k$, and so on.

\subsection{Double and triple complexes}
\label{sec:Double-triple}

If $\CCC^{\bullet,\bullet}$ is any double complex with
\emph{commuting} differentials $d_1$ and $d_2$ we denote by
$\CCC^\bullet$ or by $\Tot^\bullet\CCC$ the associated total
simple complex, with $\CCC^k = \oplus_p \CCC^{p,k-p}$, and total
differential $d(x)=d_1(x) + (-1)^p d_2(x)$, for
$x\in\CCC^{p,k-p}$.

Unfortunately (or more interestingly), we will have to consider
also \emph{triple} complexes. If $\CCC^{p,q,r}$ is such a
tricomplex with differentials $d_1,d_2,d_3$, then the associated
total complex has a total differential $d$ equal to $d=d_1 +
(-1)^p d_2 + (-1)^{p+q}d_3$ when acting on the homogeneous
component of (triple) degree $(p,q,r)$.

\subsection{Deligne complexes}
\label{sec:Deligne-complexes}

The use of Deligne cohomology is by now fairly common, so we just
introduce the notation. A brief introduction can be found in
\cite{delfreed1999-2}, a more thorough treatment can be found in
refs.\cite{esn-vie:del,bry:loop}. Recall that $M$ has dimension
$n$. The smooth Deligne complex of length $p$ is the complex of
sheaves
\begin{equation*}
  \ZZ^\bullet_\mathcal{D}: \ZZ_M
  \overset{\imath}{\longrightarrow} \sheaf{A}_M
  \overset{d}{\longrightarrow} \sheaf{A}^1_M
  \overset{d}{\longrightarrow} \dots
  \overset{d}{\longrightarrow} \sheaf{A}^{p-1}_M
\end{equation*}
$\ZZ$ is placed in degree zero and the degree of each term
$\sheaf{A}^r_M$ in $\ZZ^\bullet_\mathcal{D}$ is $r+1$. The first
differential is just the inclusion $\imath$ of $\ZZ$ in
$\sheaf{A}_M$, while $d$ is the usual de Rham differential. The
complex is truncated to zero after degree $p$.\footnote{We do not
  use the ``Algebraic Geometers' twist'' $\ZZ(p) = (2\pi
  i)^p\ZZ$, here.} The \emph{smooth
  Deligne cohomology groups} of $M$ --- denoted by
$H^q_\mathcal{D}(M,\ZZ)$ --- are the hypercohomology groups
$\HHH^q(M,\ZZ^\bullet_\mathcal{D})$. In practice, if the cover
$\cover{U}_M$ is sufficiently fine (as we assume here) these
groups can be calculated using \v{C}ech cohomology from the
double complex $\CCC^{p,q} = \check{C}^q(\cover{U}_M,
\ZZ_\mathcal{D}^p)$, equipped with the differentials $d$ and
$\deltacheck$ and the total differential $D=d
+(-1)^r\deltacheck$.

For the length $n+1$ complex used to describe the multivalued
functionals there is in fact no truncation, so it is an augmented
de Rham complex $\ZZ\rightarrow \sheaf{A}_M^\bullet$, as noted
before. In this case $H^{n+1}_\mathcal{D}(M,\ZZ)\cong
H^n(M,\CC^*)$. (See also~\cite{esn:char}.) However, in some case
the dynamical fields themselves are cocycles in
$\CCC^{\bullet,\bullet}$ for some appropriate length, so the
general formalism may be needed.

\subsection{Jets and the variational bicomplex}
\label{sec:Vari-bicompl}

We need to recall a bit of notation concerning jet bundles and
the variational complexes of local forms. An in-depth account can
be found in~\cite{saunders:jet}. We follow the approach in
\cite{zuck:action,delfreed1999-2}.

For any manifold $U$ of dimension $n$, let $\pi : E \rightarrow
U$ be a smooth fibration, with the manifold of smooth sections
$\mathcal{S} = \Gamma (U;E)$. (Later on we will consider the case
where $U \subset M$ is an open set.)  Let
$\mathit{ev}:\mathcal{S} \times U \rightarrow E$ be the
evaluation map. By taking the infinite jet of $\mathit{ev}(\phi ,
m) = \phi(m)$ at $m\in U$ it extends to $\mathit{Ev}:\mathcal{S}
\times U \rightarrow JE$, where $JE$ is the infinite jet bundle
of $E\rightarrow U$. A tangent vector $\xi$ to $\mathcal{S}$ at
$\phi$ (a ``variation'') is a section of the vector bundle
$\phi^{-1}(TE/U)$, where $TE/U$ is the vertical bundle of the
fibration $E\rightarrow U$. By taking the infinite jet $j(\phi)$
of $\phi$ we obtain a section of the vertical bundle
$j(\phi)^{-1}(JE/U)$. The vertical bundle sequence of $JE$
splits; in particular the fiber of supplementary (horizontal)
bundle at $m\in U$ is the image of $T_mM$ under $j_m(\phi)_*$. By
duality, there is a corresponding splitting of the complex of
differential forms on $JE$ into vertical and horizontal
components.

The complex of differential forms on $\mathcal{S} \times U$
splits according to the product structure into components of
bidegree $(p,q)$, and the exterior differential splits
accordingly as $d_{\mathcal{S} \times U} = \var + (-1)^p d$.  In
the complex of smooth differential forms on $\mathcal{S} \times
U$ we are only interested in the subcomplex obtained as the
inverse image under $\mathit{Ev}^*$ of the complex of
differential forms on $JE$, namely the complex of \emph{local}
forms. These are the forms whose dependency on $\phi \in
\mathcal{S}$ and tangent vectors $\xi_i$ factors (locally, in
general) through some finite jet of $\phi$ and $\xi_i$ at $m\in
U$. The local forms inherit a splitting induced by the one on
$JE$ via $\mathit{Ev}^*$. It is compatible with the one induced
by product structure of $\mathcal{S}\times U$. We denote by
$A^{p,q}_\mathrm{loc}(\mathcal{S}\times U)$ the homogeneous
component of degree $(p,q)$.

In~\cite{takens1981} (see also~\cite{delfreed1999-2} for a
different proof) Takens proves
\begin{theorem}[Takens]
  \label{thm:2}
  For $p\geq 1$ the complex
  $(A^{p,\bullet}_\mathrm{loc}(\mathcal{S})\times U,d)$ is exact
  except in top degree $\abs\bullet =n+1$.
\end{theorem}
Recall that we are shifting the form degrees by $1$. It follows
that in the variational complex $q\geq1$ when the degree shift is
in effect (as there is no $\ZZ_M$ to be placed in degree zero).
Theorem~\ref{thm:2} has two very important consequences of great
relevance to us.  First, the non exactness in top degree implies
that there are locally closed forms of top-degree which are non
exact, unlike the ordinary de Rham sheaf complex. In particular,
$A^{p,n+1}_\mathrm{loc}(\mathcal{S}\times U)$ decomposes as the
direct sum of the image of $A^{p,n}_\mathrm{loc} (\mathcal{S}
\times U)$ under $d$ and a non-exact component consisting of
\emph{source forms:} these are identified in refs.
\cite{takens1981} and \cite{zuck:action} with $(p,n+1)$-forms in
the variational bicomplex whose dependency on the variation
factors through the $0$-jet \emph{only.} Source forms are nothing
but the familiar Euler-Lagrange forms from standard variational
calculus. Indeed, what is normally done after perfoming the
variation of a Lagrangian density $\omega = L \, dx^1\wedge
\dotsm \wedge dx^n$ (we drop the chart index for convenience) is
to integrate by parts to achieve an expression containing only
the $0$-jet of the variation ``up to boundary terms'', that is, a
source form plus an exact differential. As a consequence of the
foregoing discussion we have (temporarily restoring the standard
form degrees):
\begin{lemma}
  \label{lemma:uniq}
  For a lagrangian density $\omega = L \, dx^1\wedge \dotsm
  \wedge dx^n$, there exist a unique source $n$-form $a$ and a
  local $n-1$-form $\gamma$ such that $\var \omega = a +
  d\gamma$. Moreover, if we shift $\omega$ to $\omega + d\chi$,
  the source form $a$ is unchanged.
\end{lemma}
In practical terms, lemma~\ref{lemma:uniq}
means that $\gamma$ cannot be shifted by say $\tau$, while at the
same time changing $a$ into $a -d\tau$, as the latter would not
be a source form.

Finally, the other important consequence in Theorem.~\ref{thm:2}
is that the variational complex is exact in its \emph{lowest}
degree, again unlike the de Rham one.  It follows that for $f$ a
\emph{function} with $p\geq 1$variational slots, the equation
$df=0$ implies $f=0$, rather than $f=\mathit{const}$, as it would
in the ordinary de Rham complex.

\subsection{Homology of dynamical fields and gluing of
  variational complexes}
\label{sec:Dynamical-fields}

The definition~\ref{def:1} of multivalued action functional was
given under the assumption that the components of $\Omega$ be
local forms depending on the restriction of a section of $E$ over
$M$ to the various open sets of $\cover{U}_M$. This is one of the
main examples. If $E\rightarrow M$ is a smooth fibration, a
global section $\phi$ is then a collection $\{\phi_i\}_{i\in I}$
of sections of $E\vert_{U_i}$ such that $\phi_i = \phi_j$ over
$U_{ij}$.  All the statements, however, remain valid in a more
general context.  Rather than just sections of global fibrations,
we can allow more general objects with more relaxed gluing
properties.
\begin{example}
  \label{ex:del}
  Let us consider the case of cocycles of degree $p$ in the
  \v{C}ech resolution of some Deligne complex of the (same)
  length $p$, namely an object of the form
  \begin{equation*}
    \Phi = \phi^{(0)} + \dots + \phi^{(p-1)} + c\,,
  \end{equation*}
  with $\phi^{(j)} \in \check{C}^{p-j}(\ZZ_\mathcal{D}^j)$. (And
  $c= \phi^{(p)} \in \ZZ$.) This includes the case of connections
  in line bundles and curving structures on gerbes with abelian
  band~\cite{freedwitten:anom}. The main dynamical field will be
  the collection $\{\phi^{(0)}_i\}_{i\in I}$ while the other
  members determine the gluing law, namely
  \begin{equation*}
    \phi^{(0)}_j - \phi^{(0)}_i = \pm d \phi^{(1)}_{ij}\,,\quad
    \phi^{(1)}_{ij} - \phi^{(1)}_{ik} +\phi^{(1)}_{ij} = \pm
    d\phi^{(2)}_{ijk}\,,\quad \dots
  \end{equation*}
  and so on according to the cocycle condition and the relevant
  sign rules. In a case like this, we will demand that the
  \emph{variations glue,} namely that $\var \phi^{(0)}_j =\var
  \phi^{(0)}_i$, which intuitively amounts to say that the gluing
  objects $\phi^{(1)}, \phi^{(2)}, \dots$ are spectators from the
  point of view of the dynamics.
\end{example}
Other examples we want to consider include the following.
\begin{example}
  The category of connections on a principal $G$-bundle over $M$
  for a non-abelian group $G$ with Lie algebra $\g$. The
  dynamical field is a $0$-cochain $\{A_i\}_{i\in I}$ with values
  in $\sheaf{A}_M^1\otimes \g$ with the gluing law:
  \begin{equation*}
    A_j - \ad(g_{ij}^{-1})(A_i) = g_{ij}^{-1} dg_{ij}
  \end{equation*}
  for a $1$-cocycle $\{g_{ij}\}$ with values in $G$.
\end{example}
\begin{example}
  $(G,X)$-structures: consider an action $G \times X \rightarrow
  X$ and fibrations $X\rightarrow M$ and $G\rightarrow M$, with
  corresponding sheaves of sections $\sheaf{X}_M$ and
  $\sheaf{G}_M$. The dynamical field is a collection of local
  sections $x_i$ over $U_i$ of $\sheaf{X}_M$, with elements
  $g_{ij}\in \sheaf{G}_M(U_{ij})$ acting as gluing morphisms.
\end{example}

With these examples as main motivation, we make the following
\begin{assumption-definition}\label{def:assumption}
  Let $\cover{U}_M$ be a covering of $M$, and let $\sheaf{E}_M$ a
  sheaf over $M$ with an appropriate structure, for example the
  sheaf of sections of smooth fibration $E\rightarrow M$. Let
  $\{\phi_i \in \sheaf{E}_M(U_i)\}_{i\in I}$ be a collection of
  sections. Assume that either:
  \begin{enumerate}
  \item $\sheaf{E}_M$ can be realized as the highest degree object
    in a complex of abelian groups\,;
    \begin{equation*}
      \dots \longrightarrow \sheaf{E}_M^{-2} \longrightarrow
      \sheaf{E}_M^{-1} \longrightarrow \sheaf{E}_M^0\equiv
      \sheaf{E}_M
    \end{equation*}
    and $\{\phi_i\}_{i\in I}$ completes to a cocycle of the
    appropriate length;
  \item or $\sheaf{E}_M$ can be realized as the zero level of a
    truncated simplicial object of length one:
    \begin{equation*}
      \protect\sheaf{E}_M^{-1} \rightrightarrows
      \protect\sheaf{E}_M^0\equiv \protect\sheaf{E}_M\,.
    \end{equation*}
    $\sheaf{E}_M^{-1}$ acts on $\sheaf{E}_M$ by isomorphisms.
    In this case we assume $\{\phi_i\}_{i\in I}$ is the object
    part of an appropriate decomposition of
    $\sheaf{E}_M^\bullet$, namely there are isomorphisms
    $\psi_{ij}$ over $U_{ij}$ such that $\phi_i =
    \psi_{ij}(\phi_j)$, where the $\psi_{ij}$ do \emph{not}
    necessarily satisfy a cocycle condition,
    see~\cite{breen:asterisque} for more details.
  \end{enumerate}
  In both cases, we assume that the relevant sheaves of jets
  satisfy the descent condition, namely elements of
  $\sheaf{E}_M^{-1}(U_{ij})$ induce isomorphisms $j\psi_{ij} :
  J\sheaf{E}_M\vert_{U_{ij}}\rightarrow
  J\sheaf{E}_M\vert_{U_{ij}}$ satisfying the usual compatibility
  condition over $U_{ijk}$.
\end{assumption-definition}
As a consequence we have a ``gluing lemma'' for the variational
bicomplexes above the various members of the cover $\cover{U}_M$.
Indeed, let $A^{p,q}_\mathrm{loc}(\mathcal{S}_i\times U_i)$ be
the variational bicomplex determined by
$\sheaf{E}_M\vert_{U_i}$.\footnote{Here $\mathcal{S}_i$ and
  $\sheaf{E}_M\vert_{U_i}$ are really two names for the same
  object.} Restriction to $U_{ij}$ and the action of $\psi_{ij}$
determine maps $A^{p,q}_\mathrm{loc}(\mathcal{S}_j\times U_{ij})
\rightarrow A^{p,q}_\mathrm{loc}(\mathcal{S}_i\times
U_{ij})$. Then we have
\begin{lemma}[Gluing Lemma]\label{lemma:gluing}
  For $p\geq 1$ the variational complexes
  $A^{p,\bullet}_\mathrm{loc}(\mathcal{S}_i\times U_{i})$ descend
  to a global object on $M$. Also, $\var \deltacheck =
  \deltacheck \var$.
\end{lemma}
\begin{proof}
  Essentially immediate. First of all, in general it is easy to
  verify that an isomorphism of $\sheaf{E}$ into $\sheaf{F}$
  covering the identity ``prolongs''~\cite{saunders:jet} to an
  isomorphisms between $J\sheaf{E}$ and $J\sheaf{F}$ that
  preserves the splittings in the respective complexes of
  differential forms. Thus, given the isomorphism $\psi_{ij}$
  over $U_{ij}$, the induced $j\psi_{ij}$ preserves the splitting
  in the complex of differential forms over $JE\vert_{U_{ij}}$,
  and this is consistent thanks to the descent condition
  assumption in~\ref{def:assumption}.
\end{proof}
An immediate corollary is that given a cocycle $\Phi$ of the type 
specified in \ref{def:assumption}, the variation $\var\Phi$ is a
well defined global object on $M$.

\section{Proofs}
\label{sec:Proofs}

We will (implicitly) work with the triple complex $\CCC^{p,q,r}=
\check{C}^r(\mathcal{A}^{p,q})$, where $p$ is the variational
degree, $q$ the Deligne complex one, and $r$ is the \v{C}ech
degree. The respective differentials are $\var$, $d$, and
$\deltacheck$, assumed to commute with one another. The relevant
associated total differentials are constructed according to the
rules spelled out in subsect.~\ref{sec:Double-triple}. So, for
example, we have $D=d + (-1)^q\deltacheck$, and $\triangle = \var
+(-1)^p d + (-1)^{p+q} \deltacheck$.

\subsection{Proof of Theorem~\ref{thm:1}}
\label{sec:Proof-Theorem}

Let $\Omega = \sum_{r=0}^n \omega^{(r)} + c$ be a Lagrangian
cocycle, so that $D\Omega = 0$, the latter relation being
equivalent to~\eqref{eq:2}. Assume that on $U_i$ the variation of
the Lagrangian densities $\omega^{(0)}_i$ be given by
eqn.~\eqref{eq:3}.  {}From the component of highest form degree
in~\eqref{eq:2} we have on $U_{ij} = U_i\cap U_j$,
\begin{equation}
  \label{eq:11}
  \omega^{(0)}_j -\omega^{(0)}_i = (-1)^n d\omega^{(1)}_{ij}\,,
\end{equation}
and taking the variation
\begin{displaymath}
  \var\omega^{(0)}_j -\var\omega^{(0)}_i = (-1)^n
  d\var\omega^{(1)}_{ij}\,.
\end{displaymath}
As remarked after the statement of Takens'
theorem~\cite{takens1981}, the sum in the expression for the
variation $\var\omega^{(0)}_i$ from equation~\eqref{eq:3} is a
direct one, namely the decomposition into $\mathit{source} +
\mathit{exact}$ is unique. Since equation~\eqref{eq:11} can be
interpreted as a shift in the Lagrangian density by an exact
differential, by lemma~\ref{lemma:uniq}, plus the gluing
lemma~\ref{lemma:gluing} above, we obtain
\begin{equation}
  \label{eq:7}
  a^{(0)}_j = a^{(0)}_i\,,\quad \text{and} \quad
  d\gamma^{(0)}_j -d\gamma^{(0)}_i = (-1)^n
  d\var\omega^{(1)}_{ij}\,.
\end{equation}
The second relation is an equality between differentials of
$(n-1)$-forms (i.e. of degree $n$ in the Deligne complex).
But now the variational complex is exact, so we obtain
\begin{equation*}
  \var\omega^{(1)} = (-1)^n\deltacheck \gamma^{(0)} +
  d\gamma^{(1)}\,, 
\end{equation*}
for a $\gamma^{(1)}$ of degree $(1,n-1,1)$.

We continue by recursion. For $1\leq r\leq n-1$, assume
\begin{equation}
  \label{eq:8}
  \var \omega^{(r)} = (-1)^{n-r+1} \deltacheck \gamma^{(r-1)} +
  d\gamma^{(r)}\,,
\end{equation}
then apply $\var$ to 
\begin{equation*}
  d\omega^{(r+1)} = (-1)^{n-r}\deltacheck \omega^{(r)}\,.
\end{equation*}
We have
\begin{equation*}
  d\var \omega^{(r+1)} = (-1)^{n-r}\deltacheck \var \omega^{(r)}
  = (-1)^{n-r} d\var\gamma^{(r)}\,,
\end{equation*}
having used~\eqref{eq:8}, and by Takens, again, we obtain
\begin{equation*}
  \var \omega^{(r+1)} = (-1)^{n-r} \deltacheck \gamma^{(r)} +
  d\gamma^{(r+1)}\,,
\end{equation*}
for a $\gamma^{(r+1)}$ of degree $(1,n-r-1,r+1)$, as wanted.

The last step we need to check is the relation~\eqref{eq:2} for
$r=n-1$. We have
\begin{equation*}
  d\omega^{(n)} = -\deltacheck \omega^{(n-1)}\,,
\end{equation*}
and applying $\var$ to both sides we get
\begin{equation*}
  d\var \omega^{(n)} = -\deltacheck \var \omega^{(n-1)}
  = -d \deltacheck \gamma^{(n-1)}\,,
\end{equation*}
having used~\eqref{eq:8} for $r=n-1$. Notice that the latter is
an equation for differentials of \emph{functions.} Recall our
second observation on the consequences of Theorem~\ref{thm:2}.
Thus another application of the variational complex acyclicity
property yields
\begin{equation}
  \label{eq:9}
  \var \omega^{(n)} = \deltacheck \gamma^{(n-1)}\,,
\end{equation}
without constant terms, and~\ref{thm:1} is proved.

\subsection{Proofs for the universal current}
\label{sec:Proofs-univ-curr}

The universal current calculation in proposition.~\ref{prop:1} is 
a formal manipulation of differentials. Let $\Theta = \var
\Gamma$ so that $\Theta = \sum_{r=0}^{n-1}\theta^{(r)}$, where of 
course $\theta^{(r)} = \var \gamma^{(r)}$. First, we obviously
have $\var \theta^{(r)} =0$, and, second:
\begin{equation}
  \label{eq:10}
  d\theta^{(0)} = \var d\gamma^{(0)} = -\var a^{(0)}\,,
\end{equation}
having used~\eqref{eq:3} once again. It follows
from~\eqref{eq:10} that $d\theta^{(0)}$ is in fact a well defined 
form on $M$. For the other terms, we have
\begin{align*}
  d \theta^{(r)} &= d \var \gamma^{(r)} = \var d \gamma^{(r)}\\
  &= \var \bigl( \var\omega^{(r)} - (-1)^{n-r+1}
  \deltacheck \gamma^{(r-1)} \bigr)
\end{align*}
and using~\eqref{eq:8} we obtain
\begin{math}
  d \theta^{(r)} = - (-1)^{n-r+1} \deltacheck\theta^{(r-1)}\,.
\end{math}
It follows that
\begin{equation*}
  D\Theta = \sum_{r=0}^{n-1} \bigl( d\theta^{(r)} +(-1)^{n-r}
  \deltacheck \theta^{(r)} \bigr) = -d \theta^{(0)} = \var
  a^{(0)}\,,
\end{equation*}
thanks to the last relation, which proves the proposition.

It is easy to convince oneself that all the previous
manipulations amount to the following simple calculation:
\begin{equation*}
  D \Theta = \var D \Gamma = \var ( \var \Omega - a^{(0)}) =
  -\var a^{(0)} \,.
\end{equation*}
Furthermore, $\triangle \Theta = -\var a^{(0)}$, so $\Theta$ is
indeed globally closed on $\mathcal{M}$.

\section{Conclusions and outlook}
\label{sec:Conclusions}

We have shown that under certain conditions (specified
in~\ref{def:assumption}) action functionals that appear to be
ill-defined under coordinate changes nevertheless yield a well
defined variational equation. Besides the more historical and
consolidated example of actions containing topological terms, a
compelling motivation is provided by action functionals, not
necessarily of topological flavor, arising from geometric
structures, such as those in
refs.~\cite{zogtak1987-2,aldtak2000}. In fact, the analysis
performed in ref.~\cite{aldtak2000} was one of our main
motivations in writing this note.

One limitation of the present approach lies precisely in the
assumption~\ref{def:assumption} used above. As an example, in
ref.~\cite{aldtak2000} the allowed variations were just those
that do not change the complex structure, namely the elements of
the vertical bundle along the Earle-Eells fibration over the
Teichm\"uller space. But in a geometric situation, like the one
provided by a functional depending on a complex structure, one
would like to do exactly what is beyond the scope
of~\ref{def:assumption}: performing a full variation where the
local isomorphisms themselves become dynamical fields. It is,
however, rather easy to formulate counterexamples to
lemma~\ref{lemma:gluing} and Thm~\ref{thm:1} in this more general
framework, whereby the source forms can be shown \emph{not} to
glue. One is led to conjecture that a form of Thm~\ref{thm:1}
should be valid with a cocycle $a$ of length equal to the number
of members in the complexes in assumption~\ref{def:assumption}
that are allowed to have dynamical fields. We hope to return to
this question in a future publication.

\bibliography{general}
\bibliographystyle{hamsplain}
\end{document}